\documentstyle[spie,epsfig,latexsym]{article} 
 
\title{In search for the vortex charge and the Cooper pair mass} 
 
\author{Todor M. Mishonov 
  \skiplinehalf  Laboratorium voor Vaste-Stoffysica en Magnetisme, 
  Katholieke Universiteit Leuven,\\ 
  Celestijnenlaan 200 D, B-3001 Leuven, Belgium
} 
 
\authorinfo{Phone: (++32 16) 327 193, Fax: (++32 16) 327 983, 
  E-mail:~todor.mishonov@fys.kuleuven.ac.be \\ 
  On leave from: Department of Theoretical Physics, Faculty of Physics, 
  University of Sofia,\\ 5 J. Bourchier Blvd., 1164 Sofia, Bulgaria
} 

\begin{document} 
\maketitle 
 
\begin{abstract} 
A novel experiment for determination of the charge related to 
vortices in thin superconducting film is proposed and a number of 
related experimental set-ups are also theoretically considered. 
The methods are based on the Torricelli-Bernoulli effect in 
superconductors and the phenomenology of the effect is briefly 
discussed. 
The vortex charge is expressed via the effective mass of the 
Cooper pairs, thus both parameters, inaccessible by now,  could be 
simultaneously determined. The experiment would require layered 
metal-insulator-superconductor structures and standard electronics 
employed in kinetic measurements. The quality of the 
insulator-superconductor interface should be high enough as to 
allow for observation of electric field effects similar to those 
investigated in superconducting field-effect transistors. 
The development of layer-by-layer growth technology of oxide 
superconductors provides unique possibility for investigation of 
new fundamental effects in these materials. In particular, the 
structures necessary  for determination of the  vortex charge 
could be used to study the superconducting surface Hall effect, 
Bernoulli effect, the superfluid density, etc.. In conclusion, the 
systematic investigation of new effects in oxide superconductors 
is envisaged as an important part of the material science 
underlying the oxide electronics. 
\end{abstract} 
 
 

 
\section{INTRODUCTION} 
\label{sect:intro} 

The sign change of the Hall effect observed in 
the superconducting state of many high-$T_c$ superconductors is 
one of the most puzzling problems in the electrodynamics of these 
materials~\cite{Nagaoka,Danna}. One may ask then what is the 
doping dependence\cite{Nagaoka} of this Hall anomaly and how the 
vortex-lattice melting~\cite{Danna} affects the Hall behavior? 
Alas, due to the complexity of the vortex matter many related 
problems are still not answered satisfactory if at all. It is 
quite possible that the sign reversal of the temperature 
dependence of the Hall effect could be closely related to charging 
of the vortices~\cite{Khomskii,Blatter,Hayashi,Kato}. There is no 
doubt~\cite{Hayashi} that the experimental solution of this enigma 
would provide the key towards understanding the various 
electromagnetic phenomena. On the other hand, the currently 
existing theoretical models often lead to conflicting results thus 
making it difficult to discriminate between all those competing 
explanations. In such a situation we feel it appealing to 
accelerate the selection by looking for simplicity in experiments 
with artificial structures where many of the complications typical 
for the real systems are avoided. 
 
The aim of the present paper is to propose an experiment for 
determination of the vortex charge employing transport measurement 
in a layered metal-insulator-superconductor (MIS) system. We shall 
require that the quality of the insulator-superconductor interface 
be extremely high and the insulator layer be very thin. Such a 
layered MIS structure  incorporating a high-$T_c$ film can be 
manufactured by the contemporary technology of atomic-level 
engineering of superconducting oxide multilayers and 
superlattices~\cite{technology}. In fact, structures of the kind 
are now being in use for purposes of the fundamental 
research\cite{basic} in the physics of high-$T_c$ superconductors, 
therefore the vortex charge problem can find its solution  thanks 
to the technological progress. 
The simplest possible idea behind the search for the vortex charge 
is to study the electrostatic effect related to charged 
vortices~\cite{Khomskii,Blatter}, which is analogous to 
electrostatic effects originating, in turn, in the Bernoulli 
effect~\cite{Bernoulli} due to circulating currents in a thin 
superconducting film in a vortex-free state; the numerical value 
of the angular momentum $\oint \left(m^*{\bf v}+e^*{\bf 
A}\right)\cdot {\rm d} {\bf r}$ is irrelevant for the 
current-induced contact-potential difference. The paper is 
organized as follow: in Sec.~\ref{sect:model} we derive the 
formula for the vortex charge $q_v$ expressed via the effective 
mass of Cooper pairs $m^*$ as well as the expression for the 
interface Hall conductivity $\sigma_{xy}.$ In 
Sec.~\ref{sect:set-up} we will analyze our proposed experimental 
set-up for determination of the vortex charge by measuring the 
Hall resistance of the vortex charge currents. An overview is made 
in Sec.~\ref{sect:effective mass} of different experimental 
methods for determination of the Cooper pair mass: the surface 
Hall current~\cite{surface_Hall}, subsection~\ref{sect:surface 
Hall}; the Bernoulli effect~\cite{Bernoulli}, 
subsection~\ref{sect:CPD}, and the electrostatic charge 
modulation\cite{Comment}, subsection~\ref{sect:charge modulation}. 
It is finally concluded in Sec.~\ref{sect:conclusions} that the 
vortex charge $q_v$ and the effective mass $m^*$ of fluctuation 
Cooper pairs fall into the class of the last unresolved problems 
in the physics of superconductivity. These important parameters 
enter the theories of a number of phenomena related to 
electrodynamics of superconductors and can be simultaneously 
determined by standard electronic measurements. Contemporary 
layer-by-layer growth of layered oxide structures gives the unique 
chance for finding $q_v$ and $m^*$ for high-$T_c$ materials but 
some of the proposed experiments can be realized also in MIS structures 
with conventional superconductors. 
 
\section{MODEL}
\label{sect:model} 
\subsection{Type-II superconductors} 
 
This section gives an account of the vortex charging due to the 
Bernoulli effect within the framework of London electrodynamics. 
For a superconductor in thermodynamic equilibrium the 
electrochemical potential $\zeta$ is constant and the space 
distribution of the electric potential $\varphi({\bf r})$ is 
determined by the Bernoulli-Torricelli theorem 
\begin{equation} 
  {1\over2} m^* v^2({\bf r})n(T) + \rho_{\rm tot}\varphi({\bf r}) 
    = \rho_{\rm tot}\zeta. 
  \label{eq:bernoulli} 
\end{equation} 
Formally, this equation can be derived within the framework of the 
BCS theory using the statistical mechanics methods, but its 
physical meaning is very simple---it is a consequence of the 
energy conservation. We shall further stick to the standard 
notations for the effective mass of Cooper pairs in the $ab$-plane 
$m^*,$ the superfluid velocity ${\bf v}$ related to the current 
density ${\bf j}=e^*n(T){\bf v},$ the mass density of the 
superfluid $m^*n(T),$ and the total charge of the conduction band 
$\rho_{\rm tot}=e^*n(T=0);$ at zero temperature all charge 
carriers are superfluid and according to the BCS theory 
$|e^*|=2|e|.$ The temperature dependence of $n(T)$ can be 
extracted from that of the London penetration depth for screening 
currents flowing in the CuO$_2$ plane, 
\begin{equation} 
  {1 \over \lambda^2(T)} = {\mu_0 n(T) e^{*2} \over m^*}, 
  \label{eq:depth} 
\end{equation} 
where the use of SI units is implied, $\mu_0=4\pi\times 10^{-7}.$ 
Although the temperature dependence of the superfluid ratio 
\begin{equation} 
  \frac{n(T)}{n(0)}=\frac{\lambda^2(0)}{\lambda^2(T)} 
  \label{eq:ratio} 
\end{equation} 
is related to the gap anisotropy, the hydrodynamic relation 
Eq.~(\ref{eq:bernoulli}) remains invariant. 
 
Consider now a thin cuprate film thread by a perpendicular 
magnetic field ${\bf B}=B_z \hat{\bf z}.$ As a first step we 
determine the distribution of the electric potential as a function 
of the distance to the vortex line $r=\sqrt{x^2+y^2}.$ For $r$ 
larger than the Ginzburg-Landau (GL) coherence length in the 
$ab$-plane but smaller than the penetration depth, 
  $\xi_{ab}(T) \ll r \ll \lambda_{ab}(T)$ 
one can use the Bohr-Zommerfeld relation 
\begin{equation} 
  r m^*v = \hbar. 
  \label{eq:bohr-zommerfeld} 
\end{equation} 
Substituting $v(r)=\hbar/m^*r$  from the above equation into the 
Bernoulli theorem Eq.~(\ref{eq:bernoulli}) we derive  the 
current-induced change of the electric potential 
\begin{equation} 
  \varphi(r)= -\frac{\hbar^2}{2e^*m^*} 
              \frac{n(T)}{n(0)}\frac{1}{r^2}. 
  \label{eq:potential} 
\end{equation} 
This equation is applicable not only to the volume of the 
superconductor $z<0$ but even to the superconducting surface $z=0$ 
which is supposed to be clean enough as well as to expose the 
properties of the bulk material. The superconductor is capped by a 
thin insulating layer of thickness much smaller than the 
penetration depth, $d_{\rm ins}\ll \lambda_{ab}(0).$ On top of the 
latter a thin-normal-metal layer is evaporated, hence a plane 
capacitor configuration is achieved, being in fact realized as a 
metal-insulator-superconductor (MIS) layered structure. For 
definiteness the electric potential of the normal plate is set to 
zero. Far from the vortex core, for $r>d_{\rm ins},$ the electric 
field $E_z$ of such a  plane capacitor can be considered as being 
homogeneous, 
\begin{equation} 
  E_z = \frac{\varphi}{d_{\rm ins}} = 
        -\frac{q^{(2D)}}{\epsilon_0\epsilon_{\rm ins}}, 
  \label{eq:electric} 
\end{equation} 
which is employed to express the induced on the normal plate 
surface charge density $q^{\rm (2D)}(r)$ via the Bernoulli 
potential $\varphi(r),$ 
\begin{equation} 
  q^{\rm (2D)} = \frac{\hbar^2}{2e^*m^*} 
              \frac{\epsilon_0\epsilon_{\rm ins}}{d_{\rm ins}} 
              \frac{n(T)}{n(0)}\frac{1}{r^2}, 
  \label{eq:charge_density} 
\end{equation} 
where $\epsilon_0=1/\mu_0c^2,$ $c$ being the speed of light, and 
$\epsilon_{\rm ins}$ is the relative dielectric constant of the 
insulator. We notice that $q^{\rm (2D)}(r)$ has the same sign as 
the charge of the Cooper pairs in the superconductor. On the other 
hand, the Bernoulli potential keeps the Cooper pairs on a circular 
orbits inside the vortex. The radial electric force is then equal 
to the centrifugal force 
\begin{equation} 
\label{eq:force} e^*\frac{\partial \varphi}{\partial r} = m^*\frac{v^2}{r}. 
\end{equation} 
 
The electric potential attracts the Cooper pairs and the charges 
with the same sign on the normal plate of the plane capacitor. In 
order to derive the total charge related to the vortex we have to 
integrate the charge density up to some maximum radius, 
\begin{equation} 
  \label{eq:radius} 
  r_{\rm max} = \min \left(\lambda(T),\; 
                \sqrt{\Phi_0\over B} \right), 
\end{equation} 
corresponding to the screening length $\lambda(T)$ or the typical 
intervortex distance in case of high area density of vortex lines 
$n_v=B/\Phi_0,$ where $\Phi_0=2\pi\hbar/|e^*|=2.07$~fTm$^2$ is the 
flux quantum. Supposing that the insulator layer is thin enough, 
$d_{\rm ins}\ll r_{\rm max},$ the integration of the surface 
density gives for the total vortex charge 
\begin{equation} 
  \label{eq:vortexcharge} 
  q_v=\int_{d_{\rm ins}}^{r_{\rm max}}q^{\rm (2D)}(r) 
  \;{\rm d}(\pi r^2)\approx \frac{\pi\hbar^2}{e^*m^*} 
  \frac{\epsilon_0\epsilon_{\rm ins}}{d_{\rm ins}} 
  \frac{n(T)}{n(0)}\ln\frac{r_{\rm max}}{d_{\rm ins}} = \frac{{\rm 
sign}(e^*)|e|}{8} \frac{a_0\epsilon_{\rm ins}}{d_{\rm ins}} 
  \frac{m_0}{m^*} \frac{\lambda_{ab}^2(0)}{\lambda_{ab}^2(T)}\ln 
  \kappa_{\rm eff}, 
\end{equation} 
where 
$ 
a_0=4\pi\epsilon_0\hbar^2/e^2m_0=53\;{\rm pm} 
$ 
is the Bohr radius, $m_0=9.11\times 10^{-31}$~kg is the mass of a 
free electron, and $\kappa_{\rm eff}=r_{\rm max}/d_{\rm ins}$ is a 
quantity analogous to the Ginzburg-Landau parameter 
$\kappa=\lambda_{ab}(0)/\xi_{ab}(0).$ According to our model the 
charge related to vortices is localized not in the vortex core but 
in the adjacent conducting layers: superconducting CuO$_2$ planes 
in a real high-$T_c$ crystal or the normal layer in the model MIS 
system. With this we close the electrostatic consideration of the 
vortex charge, but rhe reader is reffered to a number of ingenious 
experiments related to electrostatics of vortices which are 
suggested in Ref.~\cite{Blatter}. We believe, however, that the 
standard transport measurement have some advantage even if they 
are related to observations of pA-range and below. 
 
The next important step is to address the vortex flow regime of 
the superconducting film when a strong enough dc current density 
$j_y$ is applied through the superconducting film. This condition 
will create small dissipation and give rise to an electric field 
$E_y$ parallel to the current density. The electric field, in 
turn, creates a drift of the vortices with mean drift velocity in 
$x$-direction $v_v=E_y/B_z.$ In a coordinate system moving with 
the vortex drift velocity ${\bf v}_v$ the electric field is zero. 
We suppose that $v_v$ is much smaller than the critical depairing 
velocity $v_c=\hbar/m^*\xi_{ab}(T)$ and the Bernoulli potential is 
nearly the same as in the dissipation-free static regime. Along 
this line let us recall the fact that airplanes fly thanks to the 
Bernoulli theorem that holds true for a unviscous dissipationless 
fluid, but the significant part of the ticket price covers the 
dissipated energy. By the same token, for $v_v \ll v_c$ the 
vortex-induced charge has nearly the static value $q_v.$ Since the 
charge images will follow the vortices as shadows, the vortex flow 
will create a two dimensional (2D) current density on the surface 
of the normal metal 
\begin{equation} 
\label{normal current} 
  j^{\rm (2D)}_x = q_v n_v 
             v_v = \frac{q_v}{\Phi_0}E_y=\sigma_{xy}E_y. 
\end{equation} 
The electric field $E_y$ resides the superconducting layer, 
whereas the current $j^{\rm (2D)}_x$ exists in the normal slab. 
The 2D Hall conductivity directly gives the vortex charge 
\begin{equation} 
\label{eq:conductivity} 
  \sigma_{xy} = \frac{q_v}{\Phi_0} = \frac{q_v|e^*|}{2\pi\hbar}. 
\end{equation} 
 
For $L_x\times L_y$ rectangular shape of the MIS structure the 
voltage drop in the superconducting layer is $V_y=E_yL_y,$ the 
total current in the normal layer is $I_x=L_xj^{\rm (2D)}_x,$ and 
the interface Hall resistivity is size-independent, 
\begin{equation} 
\label{eq:Hall resistance} 
  R_{xy}\equiv\frac{V_y}{I_x}=\frac{1}{\sigma_{xy}} 
  = \frac{2\pi\hbar}{q_v|e^*|}=\frac{e^2}{q_v|e^*|}R_{\rm QHE} 
  = {1\over2}R_{\rm QHE}\frac{|e|}{q_v}, 
\end{equation} 
where $R_{\rm QHE}=2\pi\hbar/e^2= 25.813\;{\rm k}\Omega$ is the 
fundamental resistance determined by the quantum Hall effect 
(QHE). Since the vortex charge $q_v\ll |e|,$ the experiment would 
face the problem of measuring huge Hall resistances. This sets the 
first technological requirement regarding to the quality of the 
insulating layer---in order to avoid the leakage currents the 
resistance $R_{\rm MS}$ of the plane capacitor should satisfy the 
relation $R_{\rm MS}=\rho_{\rm ins}d_{\rm ins}/(L_xL_y)\gg 
R_{xy}.$ 
In the present model we used the hydrodynamic approach applicable 
for extreme type-II superconductors and completely neglected the 
influence of the geometrically small vortex core. However the 
states in vortex core can have some influence in the total charge 
of vortex core\cite{Khomskii,Blatter}. In order qualitatively to 
"interpolate" a real situation with moderate Ginzburg-Landau 
parameter $\kappa$ let us analyze the interface Hall current for a 
type I superconductor. In this case the normal "cores" are domains 
of normal metal surrounded by circulating superconducting 
currents. This problem, certainly, is only of an academic interest 
and is irrelevant for the oxide superconductors. 
 
\subsection{Interface Hall current for type-I MIS structure} 
\label{typeI} 
 
If the superconducting layer of a MIS structure is of type-I 
superconductor, in a perpendicular magnetic field $B_z$ the 
magnetic field in the normal domains is equal to the thermodynamic 
one $B_c(T)$ and is zero in the superconducting domains. The 
relative area of the normal regions is $c_N=B_z/B_c(T)$, 
correspondingly the part of the superconducting area is 
$c_S=1-B_z/B_c(T),$ thus $c_N+c_S=1$ and the external field is 
equal to the mean field $B_z=c_N B_c(T) + c_S\times 0.$ The 
contact potential difference between the normal and the 
superconducting phase (see  Eq.~(\ref{eq:bulk Bernoulli}) bellow) is 
\begin{equation} 
\label{eq:CPD} 
\varphi_N-\varphi_S= -\frac{1}{e^*n(0)} \frac{B_c^2(T)}{2\mu_0}. 
\end{equation} 
This contact potential difference creates, in turn, a difference 
in the charge density at the surface of the normal layer in front 
of the normal domain 
\begin{equation} 
\label{eq:typeI charge density} 
q^{\rm (2D)}=\frac{\epsilon_0\epsilon_{\rm ins}c_N}{d_{\rm ins}} 
\left(\varphi_S -\varphi_N\right) 
= \frac{\epsilon_0\epsilon_{\rm ins}B_c(T)}{2\mu_0d_{\rm 
ins}e^*n(0)}B_z, 
\end{equation} 
where a plane capacitor configuration is implied. 
 
When an electric field $E_y$ is applied in the superconducting 
layer the normal domains acquire a drift velocity in $x$-direction 
$v_v=E_y/B_z.$ Again, in the mobile coordinate system the domain 
structure is static and the mean electric field is zero. The extra 
charges induced in the normal layer follow the moving normal 
domains and for the 2D current $j^{\rm (2D)}_x=q^{\rm (2D)} 
v_v=\sigma_{xy}E_y$ at the surface of the normal plate we obtain 
\begin{equation} 
\label{eq:typeI sigma} 
\sigma_{xy}=\frac{\epsilon_0}{2\mu_0}\frac{B_c(T)\epsilon_{\rm ins 
}}{e^*n(0) d_{\rm ins}}=\frac{1}{R_{xy}}. 
\end{equation} 
This very small interface Hall conductivity vanishes at $T_c$ and 
its detection requires fA sensitivity. For comparison with 
Eq.~(\ref{eq:vortexcharge}) here we also give the expression for 
the induced charges per flux quantum 
\begin{equation} 
\label{eq:typeI charge} 
q_{v,I}\equiv\frac{q^{\rm (2D)}}{n_v} 
=\frac{\epsilon_0\Phi_0}{2\mu_0}\frac{B_c(T)\epsilon_{\rm 
ins}}{e^*n(0)d_{\rm ins}}. 
\end{equation} 
Of course, around every normal domain in a type-I superconductor 
$|e^*\oint {\bf A}\cdot{\rm d}{\bf r}/\hbar| \gg 1.$ 
 
Having derived the formulae, Eq.~(\ref{eq:conductivity})  and 
Eq.~(\ref{eq:typeI sigma}), concerning the new predicted effect we 
proceed with more detailed discussion and description of the 
proposed new experiment in the next section. 
 
\section{EXPERIMENTAL SET-UP FOR MEASURING THE VORTEX CHARGE} 
\label{sect:set-up} 
\begin{figure} 
  \begin{center} 
    \hspace*{0cm}\epsfig{figure=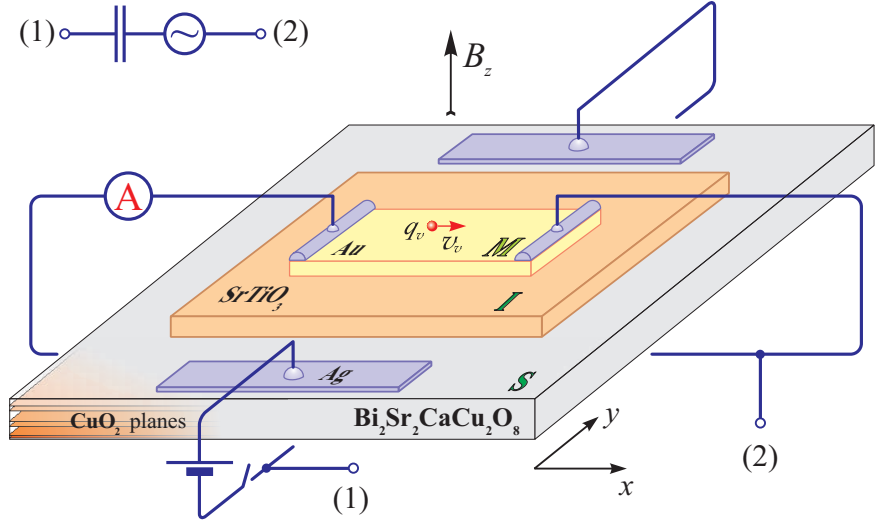,height=7cm} 
  \end{center} 
  \caption{\label{fig:setup} 
  \textit{Gedanken} set-up proposed to determine the vortex charge 
  and the Cooper pair mass. Thin Bi$_2$Sr$_2$CaCu$_2$O$_8$ layer is 
  thread by perpendicular magnetic field $B_z$. The voltage $V_y$ applied through 
  the Ag electrodes in circuit (1) creates a drift of the vortices 
  with mean velocity $v_v.$ Due to the Bernoulli effect the superfluid 
  currents around every vortex create a change in the electric 
  potential on the superconducting surface. The Bernoulli potential 
  of the vortex leads to an electric polarization on the normal Au 
  surface. The charge $q_v$, related to the vortex, has the same drift 
  velocity $v_v.$ The corresponding current $I_x$ in 
  circuit (2) can be read by a sensitive ammeter. The quality of 
  the SrTiO$_3$ plate should be high enough so as to allow detection of 
  the interface Hall current without being significant perturbed by 
  the leakage currents between circuits (1) and (2). 
} 
   \end{figure}

To begin with, we have sketched a "\textit{gedanken}" experimental 
set-up in Fig.~\ref{fig:setup}. 
%
The contemporary technology of layer-by-layer growth of oxide 
superconductors opens the possibility for realization of such a 
layered structure---a superconducting film protected by an 
insulating plate. Moreover, we consider that a MIS plane capacitor 
is one of the simplest possible systems employed in the 
fundamental research towards further technical applications. 
Therefore we believe that the suggested experiment could become a 
standard tool in studying the quality of the 
insulator-superconductor interface. In order to check whether this 
idea is another case of a science fiction or, vice versa, is a smoking gun we 
provide below a numerical example involving an acceptable set of 
parameters which have been collected from various references: 
$m^*= 11 m_0$ (Ref.\cite{Comment}), $\xi_{ab}(0)= 1.1$~nm, $d_{\rm 
ins}= 15\;$nm, $d_{\rm ins}/\epsilon_{\rm ins}=1$~nm 
(Ref.\cite{McKee}), $\lambda_{ab}(0)=150$~nm (Ref.\cite{Fiory}). For 
an illustration we take as well: $B_z=100$~mT, $E_y=1$~V/cm, and 
$L_x= L_y=1$~mm. 
 
The value of $B_z$ we chose imply for the following parameters: 
$n_v=B_z/\Phi_0= 4.83\times 10^{13}{\rm m}^{-2},$ $L_xL_yn_v= 
48\times 10^{6},$ and $1/\sqrt{n_v}=144\mbox{ nm } \simeq 
\lambda_{ab}(0).$ For a model estimate we also take $r_{\rm 
min}\approx 150$~nm. It is now straightforward to work out the 
vortex charge at liquid-helium temperature, \textit{i.e.} in the 
temperature range far below $T_c.$ In this case the substitution 
of the above mentioned set of parameters in 
Eq.~(\ref{eq:vortexcharge}) gives 
\begin{equation} 
\label{eq:charge evaluation} 
  \frac{q_v}{|e|}= \frac{1}{8}\cdot\frac{53} 
                   {1000}\cdot\frac{1}{11}\cdot\ln(10) 
                 = 1.386\times10^{-3}\simeq {1 \over 1000}. 
\end{equation} 
The so estimated  $q_v\simeq 10^{-3} |e|$ is in agreement with 
another model evaluation due to Khomskii and 
Freimuth\cite{Khomskii}. Further, Eq.~(\ref{eq:Hall resistance}) 
gives $R_{xy}=9.35$~M$\Omega$ and the electric field chosen gives 
for the voltage $V_y=100$~mV, therefore for the Hall current we have 
$I_x=R_{xy}V_y= 11$~pA. Lastly, the vortex drift velocity 
$v_v=E_y/B_z=1$~km/s, which is one order of magnitude smaller than 
the depairing velocity at $T=0,$ 
$v_c=\hbar/m^*\xi_{ab}(0)=9.6$~km/s. We note that the resistance 
of the capacitor should be thus at least $R_{\rm 
MS}=100$~M$\Omega.$ 
 
For conventional superconductors similar evaluations show that 
effect is less but still observable. One can consider, for 
example, a thin Nb metal film grown by molecular beam epitaxy, and 
an Al layer after oxidation in natural condition could give a good 
insulator layer. All technologies for planar Josephson junctions 
provide as a rule metal-insulator interface of sufficient quality. 
Only the insulator layer should be thick enough to prevent leakage 
tunneling. 
 
The example analyzed above shows that the proposed experiment is 
in principle possible to be carried out but we find it difficult 
to anticipate all problems that could arise in the course of it. 
For instance, due to a good capacitance cross-talk the noise 
created by the vortex motion in the superconducting layer will be 
transmitted to the normal layer thus disturbing the detection of 
the small Hall current. We believe, however, that similar problems 
could be surmounted, given the challenge of the novel physics 
underlying the vortex charge. Furthermore, it is quite possible 
that the charge, concentrated in the vortex core, is comparable to 
the charge outside, so only a detailed analysis within the 
microscopic theory can shine a light on the latter point. In order 
to verify whether a hydrodynamic approach based upon the Bernoulli 
effect suffices to quantitatively describe the predicted 
vortex-charge interface current one needs independent methods to 
determine the effective mass of the Cooper pairs. In the next 
section we will analyze similar experiments employing artificial 
MIS structures. 
 
\section{HOW TO MEASURE THE COOPER PAIR MASS} 
\label{sect:effective mass} 
 
Before addressing the problem of measuring the Cooper pair 
effective mass $m^*$ let us analyze a parallel between the latter 
issue and the civil engineering, where in a static approximation 
only the weight $W=mg$ is essential for a construction. In this 
approximation the masses could reach colossal values if we 
renormalize the earth acceleration $g\rightarrow 0$. The 
uncertainty, however, immediately disappears during the first 
earthquake when a dynamical problem should be solved. Just the 
same is the situation with the superconducting order parameter 
$\Psi$---in the static GL theory the superfluid density 
$n=|\Psi|^2$ and the effective mass $m^*$ are inaccessible 
separately. They are contained in the experimental parameters, 
such as the penetration depth Eq.~(\ref{eq:depth}), only via the 
ratio $n/m^*.$ In order to determine the effective mass one has to 
investigate some dynamic phenomenon, which is time-dependent. Due 
to phase invariance, however, the time $t$ could participate only 
in the gauge invariant derivative $\left(i\hbar 
\partial/\partial t - e^* \varphi\right)\Psi ,$ that is why  electric 
field effects in superconductors are to be studied. The subtle 
point is that  the latter are already dynamic effects even if the 
electric fields are static. One therefore needs to perturb the 
thermodynamic equilibrium of the superconductor as slightly as 
possible and all methods for determination of the effective mass 
$m^*$ of Cooper pairs thus become effectively ac methods, based on 
the electrostatic effects in superconductors. The set-up proposed 
to determine the vortex charge, Fig.~\ref{fig:setup}, is a MIS 
device having four terminals. Probably the most simple method to 
accomplish the task would be to use the same MIS structure without 
making any contacts on the superconducting layer and to 
investigate the surface Hall current\cite{surface_Hall} as 
described in the next subsection. 
 
\subsection{Surface Hall current} 
\label{sect:surface Hall} 
 
\begin{figure} 
  \begin{center} 
    \hspace*{0cm}\epsfig{figure=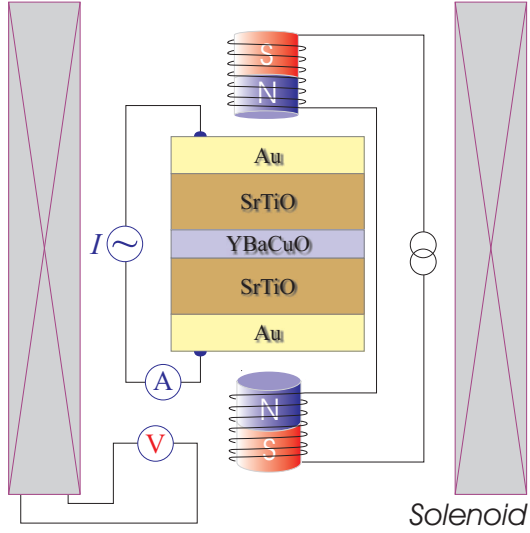,height=7cm} 
  \end{center} 
  \caption{\label{fig:surface_Hall} 
Set-up for observation of surface Hall 
current induced by a normal to the superconducting film electric 
induction $D_z$ and nearly homogeneous parallel-to-the-film 
magnetic field $B_r.$ The core ingredient is a layered MIS 
structure (see text) in the field of a plane capacitor (Corbino 
geometry; schematically, not to be scaled). The ac voltage 
generator creates current $I$ through the plane capacitor, and the 
dc current source generates opposite oriented magnetic poles in 
the drive coils and a radial magnetic field $B_r$ in the plane of 
the superconducting film. A many-turn solenoid is used to detect 
the ac magnetic moment $M_z$ of the circulating surface Hall 
currents $j_\varphi^{\rm(2D)}.$ } 
\end{figure}

This physical effect reffers to the 2D surface currents ${\bf 
j}^{\rm (2D)}$ at the surfaces of a thin ($d_{\rm film}\ll 
\lambda_{ab}(0)$) superconducting film induced by a 
normal-to-the-layer electric induction ${\bf D}_n$ and 
parallel-to-the-layer magnetic field ${\bf B}_t$ 
\begin{equation} 
  \label{eq:surface Hall} 
  {\bf j}^{\rm (2D)}=\frac{e^*}{m^*}d_{\rm 
  film}\frac{\lambda_{ab}^2(0)}{\lambda_{ab}^2(T)}\;{\bf D}_n \times 
  {\bf B}_t, 
\end{equation} 
where the Cooper pair mass $m^*$ is the material constant of the 
effect. This is an electrostatic effect and the superconducting 
film is in vortex-free state. The dissipation is zero and the 
superconductor is in thermodynamic equilibrium. A symmetric 
layered structure is grown by capping of the superconducting film 
with an insulator layer. Two normal metal layers are evaporated on 
the protecting insulator layer and on the back side of the 
substrate thus achieving a plane capacitor configuration. The 
normal-metal electrodes are circles with radius $R$ and a cartoon 
of the experimental set-up in Corbino geometry is shown in 
Fig.~\ref{fig:surface_Hall}. 
%
Exploiting the axial symmetry of the geometry Eq.~(\ref{eq:surface 
Hall}) reads as $j^{\rm (2D)}_{\varphi}\propto D_z B_r$ and for 
the total magnetic moment of the circulating currents we have 
\begin{equation} 
\label{eq:magneticmoment} 
M(t)=\int_0^R (\pi r^2)j^{\rm (2D)}_{\varphi}(r) {\rm d}r 
=\frac{e^*}{m^*}d_{\rm 
film}\frac{\lambda_{ab}^2(0)}{\lambda_{ab}^2(T)}\;D_z(t) \int_0^R 
(\pi r^2)B_r(r){\rm d}r 
\end{equation} 
 
This small magnetic moment could be difficult to detect against 
the large background due to the dc magnets creating $B_r.$ We 
derive a static magnetic moment and the next natural step is to 
consider in a quasistatic approximation the electric induction 
$D_z$ as being time-dependent, $D_z=D_z(t)$ . The ac magnetic 
moment can be detected by the electromotive voltage 
\begin{equation} 
  \label{eq:electromotive} 
  {\cal E} (t)= -\mu_0 \nu \frac{{\rm d}M(t)}{{\rm d}t}, 
\end{equation} 
induced in the solenoid having $\nu$ turns per unite length. The 
total charge of the capacitor is $(\pi R^2)D_z$ and the time 
derivative of the electric induction, 
\begin{equation} 
  \label{eq:chargingcurrent} 
  \frac{{\rm d}D_z(t)}{{\rm d}t}=\frac{I(t)}{\pi R^2} 
\end{equation} 
can be expressed by the current $I(t)$ charging the capacitor. For 
the electromotive voltage we finally obtain the equation 
\begin{equation} 
  \label{eq:finalvoltage} 
  {\cal E}(t)=R_{\rm eff} I(t)-M_{12}\frac{{\rm d}I(t)}{{\rm d}t}, 
\end{equation} 
where 
\begin{equation} 
  \label{eq:Reff} 
  R_{\rm eff}= -\mu_0\frac{e^*}{m^*}\frac{\nu d_{\rm film}}{\pi 
  R^2}\frac{\lambda_{ab}^2(0)}{\lambda_{ab}^2(T)} \int_0^R (\pi 
  r^2)B_r(r){\rm d}r 
\end{equation} 
is the effective resistance describing this new electrodynamic 
effect created by the effective mass $m^*.$ The experimental 
difficulties might be related with the careful compensation of the 
mutual inductance $M_{12}$ between the solenoid and the ac 
generator charging the MIS plane capacitor. The rigorous analysis 
of the experiment requires the knowledge of the break-through 
voltages of the MIS structure and the noise induced in the 
detecting coil, but in any case this auxiliary experiment would be 
easier to perform than the detection of vortex charge currents. 
 
In the following we will also provide an elementary derivation of 
the formula for the surface Hall current Eq.~(\ref{eq:surface 
Hall}) using the London electrodynamics. Let us trace the 
trajectory of a London {\em superconducting electron} 
(\textit{i.e.} a Cooper pair) crossing the circular 
superconducting film during the charging of the MIS plane 
capacitor, Fig.~\ref{fig:surface_Hall}. The {\em superconducting 
electron} leaves the inital surface of the film 
 with zero velocity $v_{\varphi}(t_i)=0$, experiences the 
Lorentz force while traveling across the film 
\begin{equation} 
  \label{eq:Newton} 
  m^*\frac{{\rm d}v_{\varphi}(t)}{{\rm d}t}= e^*_z B_r, 
\end{equation} 
and arrives at the opposite surface of the film at the $t_f,$ 
\textit{i.e.} 
\begin{equation} 
  \label{eq:path} 
  \int_{t_i}^{t_f} v_z(t) {\rm d} t = d_{\rm film} 
\end{equation} 
with an additional azimuthal velocity component 
\begin{equation} 
  \label{eq:finalvelocity} 
  v_{\varphi}= \frac{e^*}{m^*}d_{\rm film} B_r. 
\end{equation} 
For $T=0$ all charges are superfluid and the electric induction 
determines the surface (or 2D) excess charge density 
$D_z=e^*n^{\rm (2D)}.$ For the surface current density of these 
polarization charges we therefore have 
\begin{equation} 
\label{eq:polarizationcurrent} 
j_{\varphi}^{\rm (2D)}= e^*n^{\rm (2D)}v_{\varphi}=D_z v_{\varphi} 
= \frac{e^*}{m^*}d_{\rm film} D_z B_r. 
\end{equation} 
For non-zero temperatures one has to take into account the thermal 
dissociation of the {\em superconducting electrons}, 
$e^*\rightarrow e+e,$ and the appearance of a normal fluid. Thus, 
taking into account the superfluid part, 
\begin{equation} 
\label{eq:superfluid ratio} 
j_{\varphi}^{\rm (2D)}(T>0)= \frac{n(T)}{n(T=0)} 
j_{\varphi}^{\rm(2D)}(T=0) 
\end{equation} 
we recover the basic equation Eq.~(\ref{eq:surface Hall}). The BCS 
treatment certainly gives the same result because the London 
electrodynamics is not a mere, naive phenomenological alternative 
to the microscopic BCS theory, instead it should be viewed as an 
efficient tool to apply the BCS theory to low frequencies 
$\omega\ll \Delta/\hbar$ and small wave-vectors $k\xi_{ab}(0).$ 
 
Analogous experiment could be performed with a bulk crystal or 
thick film $d_{\rm film}\gg \lambda_{ab}(0).$ In this case in the 
initial Eq.~(\ref{eq:surface Hall}) and the final result, 
Eq.~(\ref{eq:Reff}), the thickness of the film $d_{\rm film}$ 
should be replaced with the penetration depth $\lambda_{ab}(T)$ 
and the formula for the surface current then reads as 
\begin{equation} 
  \label{eq:surface Hall bulk crystal} 
  {\bf j}^{\rm 
  (2D)}=\frac{e^*}{m^*}\frac{\lambda_{ab}^2(0)}{\lambda_{ab}(T)}\;{\bf 
  D}_n \times {\bf B}_t. 
\end{equation} 
The investigation of the temperature dependence of this effect can 
give a new method for determination of the temperature dependence 
of the penetration depth $\lambda_{ab}(T).$ A SrTiO$_3$ layer 
should be grown on the fresh cleaved surface of 
Bi$_2$Sr$_2$CaCu$_2$O$_8$ crystal and a circular Au electrode 
needs to be overgrown on the protecting layer. One plate of the 
capacitor is the bulk high-$T_c$ crystal and the other one is the 
Au layer. In order to avoid frozen vortices the constant magnetic 
field of the dc drive coil must be applied after cooling down to 
low temperatures. An ac voltage should be applied to the plane 
capacitor, a lock-in ammeter will measure the polarization 
current, and the induced due to the effect ac magnetic moment can 
be detected by a lock-in voltmeter connected to the detector coil. 
For derivation of the above formula Eq.~(\ref{eq:surface Hall bulk 
crystal}) 
 we have to use: (i) the distribution of the 
vector-potential at depth $|z|$ in the superconductor and some 
fixed radius $r,$ 
\begin{equation} 
\label{eq:Ochsenfeld} 
A_{\varphi}(z)=B_r\lambda_{ab}(T) \exp \left( 
-\frac{|z|}{\lambda_{ab}(T)}\right), 
\end{equation} 
where $B_r(r)$ is the tangential magnetic field at the 
superconducting surface; (ii) the London-BCS formula for the 
current response of the superconductor (the polarization 
operator), 
\begin{equation} 
\label{eq:polarizationoperator} 
j_{\varphi}= -\frac{A_{\varphi}}{\mu_0\lambda^2_{ab}(T)}, 
\end{equation} 
and (iii) the formula for the bulk (3D) density of the superfluid 
polarization charges 
\begin{equation} 
\label{eq:bulk superfluid} 
e^*n(z)= D_z 
\frac{\lambda_{ab}^2(0)}{\lambda_{ab}^2(T)}\;\delta(z), 
\end{equation} 
where $\delta$ stands for the Dirac $\delta$-function. The 
effective mass $m^*$ can be determined not only by the surface 
Hall effect but also from the Bernoulli effect for which the BCS 
theory was developed by Omel'yanchuk and 
Beloborod'ko~\cite{Omelyanchuk} as well as from all other 
predictions of the London theory. The existence of Bernoulli 
effect for conventional superconductors is experimentally 
confirmed; some references can be found, for example, in 
Ref.~\cite{Bernoulli}. In the next subsection we give a brief 
account of the suggested here Cooper pair mass spectroscopy. 
 
\subsection{Bernoulli effect in thin superconducting film} 
\label{sect:CPD} 
 
\begin{figure} 
  \begin{center} 
    \hspace*{0cm}\epsfig{figure=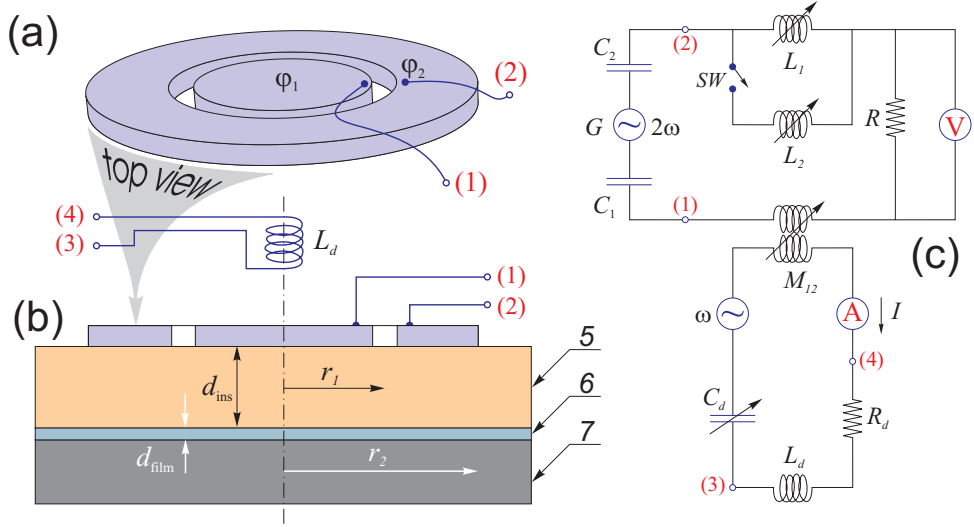,height=7cm} 
  \end{center} 
  \caption{\label{fig:bernoulli ac} 
Cooper pair mass spectroscopy based on the Bernoulli potential 
(after Ref.~\cite{Bernoulli}). (a) top view (b) cross section, (c) 
equivalent electric scheme. Two electrodes, circle- (1) and 
ring-shaped electrode (2), should be produced on the insulating 
layer capping the superconducting film. (3) and (4) denote the 
contacts of the drive coil with inductance $L_d$ and resistance 
$R_d$; (5)---insulator layer with thickness $d_{\rm ins};$ 
(6)---superconducting film with thickness $d_{\rm film} < 
\lambda_{ab}(0)$; (7)---substrate; $M_{12}$---mutual inductance; 
$L_1,$ $L_2$---variable inductances; $R$---load resistor; 
$V$---voltmeter; $A$---ammeter; $SW$---switch; $C_d$---capacitor 
of the drive resonance contour with resonance frequency $\omega;$ 
$G$---Bernoulli voltage generator with doubled frequency 
2$\omega;$ $C_1,\; C_2$---capacitances between the superconducting 
film and metal electrodes (1) and (2). This figure and the 
underlying author's idea have been used in the discussions in 
Refs.~\protect{\cite{Blatter,Khomskii}} on the vortex charge 
problem; for distribution of the electric force lines of the 
circulating currents see Fig.~1 of Ref.~\protect{\cite{Blatter}}. 
} 
\end{figure}

The experimental set-up for a current-induced Cooper pair mass 
spectroscopy is presented in Fig.~\ref{fig:bernoulli ac}. 
%
The Bernoulli effect is related to a current-induced 
contact-potential difference that can be measured by the 
electrostatic polarization of a normal metal electrode which 
covers the surface of the superconductor, 
 forming a plane capacitor. For the averaged change of the electric potential 
beneath the electrode the Bernoulli theorem 
Eq.~(\ref{eq:bernoulli}) gives 
\begin{equation} 
\label{eq:averaged potential} 
e^* \langle \varphi \rangle = -\frac{n(T)}{n(T=0)}\; \langle 
\frac{1}{2} m^* v^2\rangle, 
\end{equation} 
\textit{i.e.} the Bernoulli potential is proportional to the averaged 
kinetic energy of Cooper pairs beneath the electrode. For thin 
films, $d_{\rm film}< \lambda_{ab}(0),$ the current across the 
layer is more or less homogeneous $j^{\rm (2D)}=d_{\rm film}j$ and 
we have to substitute in this equation $v\approx j^{\rm 
(2D)}/(d_{\rm film} e^* n(T)).$ Then the formula for the Bernoulli 
potential takes the form 
\begin{equation} 
  \label{eq:potential 2D} 
  \langle\varphi\rangle 
  = -\frac{m^*\langle(j^{\rm (2D)})^2\rangle}{2e^{*3}d_{\rm film}^2n(0)n(T)} 
  = -\frac{e^*\mu_0^2\lambda_{ab}^2(0)\lambda_{ab}^2(T)}{2m^*d_{\rm 
    film}^2}\langle(j^{\rm (2D)})^2\rangle 
  = -\frac{L_{\Box}(T)}{2e^*n(0)d_{\rm film}}\langle(j^{\rm (2D)})^2\rangle, 
\end{equation} 
where 
\begin{equation} 
  \label{eq:kinetic inductance} 
  L_{\Box}(T)\equiv\frac{m^*}{e^{*2}n(T)^{\rm (2D)}} 
  = \mu_0\frac{\lambda_{ab}^2(T)}{d_{\rm film}} 
\end{equation} 
is the kinetic inductance which can be measured directly by means 
of the mutual inductance method~\cite{Fiory,Rogers}, $n^{\rm 
(2D)}(T)=n(T)d_{\rm film}$ is the area density of Cooper pairs and 
$\langle(j^{\rm (2D)})^2\rangle$ is the averaged square of the 2D 
supercurrent beneath the electrode whose distribution has to be 
found by solving a magnetostatic problem. If two electrodes were 
grown on the superconductor surface, the Bernoulli voltage 
\begin{equation} 
\label{eq:Bernoulli voltage} 
V_{\rm Bernoulli}= \langle \varphi \rangle_2- \langle \varphi 
\rangle_1 
\end{equation} 
can be considered as a voltage generator sequentially connected to 
two capacitors $C_1$ and $C_2$ as depicted in 
Fig.~\ref{fig:bernoulli ac}~(c). The currents induced in the 
superconductor film are proportional to the current through the 
drive coil $L_d,$ $j^{\rm (2D)}\propto I_{\rm drive},$ 
Fig.~\ref{fig:bernoulli ac}~(b,c). The coefficient $A_a$ of this 
proportion $\langle \left( j^{\rm (2D)} \right) \rangle=I_{\rm 
drive}^2/A_a$ has dimension of area~\cite{Bernoulli}. According to 
Eq.~(\ref{eq:potential 2D}) an ac current $I_d\propto \cos(\omega 
t)$ will create an ac Bernoulli voltage of doubled frequency 
$V_{\rm Bernoulli}\propto \cos(2\omega t).$ Initially, in the 
switched-off regime, when the detector contour resonates at 
frequency $\omega=1/(L_1C)^{1/2},$ where $C=C_1C_2/(C_1+C_2)$ the 
parasite mutual inductance between the drive coil contour  and the 
detector contour must be carefully annulled by a small tunable 
mutual inductance $M_{12}.$ After that taking $L_2\approx L_1/3$ 
in switched-on regime the detecting contour will resonate at 
doubled frequency $2\omega=1/(LC)^{1/2},$ $L=L_1L_2/(L_1+L_2).$ In 
resonance conditions the Bernoulli voltage can be directly 
detected by a lock-in voltmeter with a low noise preamplifier. If 
we know the penetration depth $\lambda_{ab}(T)$ the measured 
Bernoulli voltage, according to the Eq.~(\ref{eq:potential 2D}), 
gives the effective mass of Cooper pairs $m^*.$ 
 
If thick films, $d_{\rm film}\gg \lambda_{ab}(0),$ or bulk single 
crystals are to be used for such experiment we have to substitute 
in Eq.~(\ref{eq:averaged potential}) the London formula for the 
velocity $m^*{\bf v}=-e^*{\bf A},$ which is a trivial consequence 
of the Newton equation $m^*{\rm d}{\bf v}/{\rm d} t= e^*{\bf E}$ 
for a nearly homogeneous electric field ${\bf E }(t)=-\partial 
{\bf A}/\partial t.$ Combining with Eq.~(\ref{eq:Ochsenfeld}) we 
obtain 
\begin{equation} 
\label{eq:bulk Bernoulli} 
\langle \varphi \rangle= - R_{\rm LH} \langle p_B \rangle,\qquad 
p_B=\frac{B_r^2}{2\mu_0}, \qquad 
R_{\rm LH}\equiv \frac{1}{e^* n(T=0)}, 
\end{equation} 
where $p_B$ is the pressure of the tangential to the 
superconducting surface magnetic field, and the $R_{\rm LH}$ is 
the temperature independent\cite{Bernoulli} London-Hall constant 
expressed via the total volume density of conduction band 
$\rho_{\rm tot}=e^*n(0)=1/R_{\rm LH}.$ We consider the Greiter, 
Wilczek and Witten's~\cite{Greiter} prediction for a temperature 
dependence of the London-Hall constant as being erroneous and the 
problem still waits for its experimental solution. For type-I 
superconductors the Eq.~(\ref{eq:bulk Bernoulli}) can be applied 
up to $B_c(T)$ obtaining in this way the contact potential 
difference Eq.~(\ref{eq:CPD}). It is still questionable whether 
the thermal-induced contact-potential difference 
\begin{equation} 
\label{eq:thermal} 
\varphi(T_2)-\varphi(T_1)= -\frac{1}{e^*n(0)} 
\frac{B_c^2(T_2)-B_c^2(T_1)}{2\mu_0} 
\end{equation} 
may be measured, but if the answer is positive this effect can 
be used to determine 
the thermodynamic critical field $B_c(T)$ even for type-II 
superconductors. In any case the fluctuation of the temperature 
should be taken into account in the experiments aiming to observe 
the Bernoulli effect. 
 
The realistic experiment proposed in Ref.~\cite{Bernoulli} can be 
substantially simplified (cf. Ref.~\cite{Simkin}): the ring 
electrode capacitor can be substituted by a short circuit, and the 
central one could cover the whole facet. We stress that at least 
one capacitive connection is indispensable. The voltmeters do not 
measure any voltage difference but just the \textit{difference in 
the electrochemical potential} (even nowadays almost 99\% of the 
experimentalists are unaware of what a voltmeter really measures)! An 
error of the kind has prevented Lewis~\cite{Lewis} during his 
pioneer investigations in the period 1953--1955 from observing the 
Bernoulli effect in superconductors soon after it has been 
predicted by London\cite{London}. Lewis did not use the capacitive 
connection but he introduced all other necessary ingredients: 
lock-in voltmeter with nV sensitivity, ac magnetic field and 
doubling of the frequency. Now it is worthwhile measuring both the 
Bernoulli effect and the surface Hall effect in the same sample. 
At known total charge density $\rho_{\rm tot},$ Eq.~(\ref{eq:bulk 
Bernoulli}), and penetration dept $\lambda_{ab}(0),$ 
Eq.~(\ref{eq:surface Hall bulk crystal}), the Cooper pair mass can 
be determined as $m^*=\mu_0 e^*\rho_{\rm tot}\lambda^2_{ab}(0).$ 
Despite the $40\times 10^3$ papers published on high-$T_c$ 
superconductivity, without the Cooper pair mass the physics of 
superconductivity remains Hamlet without the Prince, with only the 
role of Ophelia performed by \textit{onnagata}\footnote{female 
impersonator in \textit{kabuki} theater}. In the next subsection 
we briefly describe the only, to the best of our knowledge, 
reliable experiment for determination of effective mass $m^*.$ 
 
\subsection{Electric charge modulation of the kinetic inductance} 
\label{sect:charge modulation} 
 
When an electric voltage is applied to a MIS plane capacitor the 
charging of the superconducting surface will create a change of 
the 2D superfluid charge density 
\begin{equation} 
\label{eq:charge modulation} 
e^*n^{\rm (2D)}= e^*d_{\rm film}n(T)+D_z\frac{n(T)}{n(0)} 
=\left( d_{\rm film}\rho_{\rm tot} + D_z 
\right)\frac{\lambda_{ab}^2(0)}{\lambda_{ab}^2(T)}. 
\end{equation} 
It is then easily worked out from Eq.~(\ref{eq:kinetic 
inductance}) that this creates a modulation of the kinetic 
inductance and the derivative determines~\cite{Comment} the 
effective mass 
\begin{equation} 
\label{eq:modulated inductance} 
m^*= - e^* L_{\Box}(0) L_{\Box}(T)\frac{\delta D_z}{\delta 
L_{\Box}(T)}. 
\end{equation} 
This simple picture gets complicated due to $T_c$-changing upon 
electrostatic doping of the material, but below the critical 
region this experiment confirms~\cite{Reply} a temperature 
independent effective mass $m^*$. When $m^*$ and all other 
parameters of the superconductor are already determined we can 
turn to the vortex charge problem. 

\section{DISCUSSION AND CONCLUSIONS} 
\label{sect:conclusions} 

The preceding analysis demonstrates that the proposed electronic 
measurements are feasible and the suggested experimental programme 
could be soon realized. The appearance of the first good samples 
would immediately lead to the solution of the problem concerning 
the vortex charge and Cooper pair mass. These two parameters, 
$q_v$ and $m^*,$ might fall in the lime-light of the physics of 
superconductivity in the nearest future. As a by-product the 
Cooper pair mass spectroscopy could become a standard tool for 
testing the quality of the superconducting films for future 
superconductor electronics. Even in the present paper we suggested 
two or three new effects thus there is no doubt that new physics 
will emerge from the development of the layer-by-layer oxide 
technology. Let us also list some of the main results of this 
study: the formulae for vortex charge Eqs.~(\ref{eq:vortexcharge}) 
and (\ref{eq:charge evaluation}), vortex conductivity 
Eqs.~(\ref{eq:conductivity}) and (\ref{eq:Hall resistance}), 
surface Hall current for bulk crystals Eq.~(\ref{eq:surface Hall 
bulk crystal}), interface Hall conductivity for type-I 
superconductors Eq.~(\ref{eq:typeI sigma}), thermal-induced 
contact-potential difference Eq.~(\ref{eq:thermal}), etc.. 
 
Finding a solution to the vortex charge problem by employing a 
model system, where the superconducting and the polarized layers 
are separated, will immediately trigger the answer to the question 
about what is the charge induced in the adjacent CuO$_2$ layers by 
a pancake vortex. One may further ask about the fate of the charge 
cloud when the pancake vortices "polymerize" in a vortex line, and 
what is the influence of the vortex charge in the vortex-vortex 
interaction and correlation. According to our analysis of the Bernoulli 
effect the charge will concentrate at the end of vortex lines, 
at kinks and sharp turns of stacks of pancake vortices. Needless 
to say, the clear solution of some model problems is always useful 
in the search for solution to the complex problems in material 
science. 
 
The problem of determining the vortex charge by a transport 
measurement brings us back to one of the first ideas of the 
electron physics. Only two months after the discovery of the 
electron~\cite{Mott} Francis Mott made the first attempt to 
observe the influence of the electric fields and surface charges 
on the conductivity of Pt. Likewise, the vortex charge current has 
led us to another immortal idea of the XIX century---the Kelvin 
vortex model of the "atom". Starting from a hydrodynamic approach, 
we were able to realize that the hydrodynamic excitations can 
propagate as particles and that the charge related to vortex 
"atoms" gives a measurable electric current. 
 
\acknowledgments
 
The author is thankful to Prof.~Y.~Bruynseraede and 
Prof.~J.~Indekeu for the hospitality and the interest in this 
investigation. The author is very much indebted to I.~Bozovic, 
J.~Clayhold, V.~Geshkenbein, D.~I.~Khomskii, A.~I.~Larkin, 
Y.~Matsuda, M.~Simkin, K.~Temst and A.~Volodin for the 
correspondence, papers provided, comments and the clarifying 
discussions. It is a pleasure to thank E.~Penev for critically 
reading the manuscript, providing the electronic versions of the 
figures and a many suggested improvements. The author appreciates 
the collaboration with M.~Mishonov in deriving the results in 
subsection~\ref{typeI} and the indispensable help of V.~Mishonova. 
This paper was supported by the Belgian DWTC, the Flemish 
Government Programme VIS/97/01, the IUAP and the GOA. 
 
\bibliographystyle{spiebib} 

\end{document}